\documentclass[10pt,conference]{IEEEtran}
\IEEEoverridecommandlockouts

\usepackage{cite}
\usepackage{amsmath,amssymb,amsfonts}
\usepackage{algorithmic}
\usepackage{graphicx}
\usepackage{textcomp}
\usepackage{xcolor}
\usepackage[hidelinks]{hyperref}
\usepackage{multirow}

\usepackage{tcolorbox}
\newtcolorbox{rqanswer}[1]{colback=gray!8,colframe=gray!60,boxrule=0.5pt,arc=1pt,
  left=4pt,right=4pt,top=2pt,bottom=2pt,toptitle=1pt,bottomtitle=1pt,
  coltitle=black,colbacktitle=gray!25,fonttitle=\bfseries,title=#1}

\def\BibTeX{{\rm B\kern-.05em{\sc i\kern-.025em b}\kern-.08em
    T\kern-.1667em\lower.7ex\hbox{E}\kern-.125emX}}
\begin{document}

\title{Can AI Agents Help Robotic Engineers? Lessons from Two Focus Days in Industry
\thanks{This work was carried out as part of the PLC-GPT project, a collaboration between the University of Passau and AKE Technologies funded by the Bavarian State Ministry of Science and Arts. We are grateful to AKE Technologies for granting us access to their data and for the feedback their engineers provided regarding the utility that guided the use case development and assessment.}
}

\author{\IEEEauthorblockN{1\textsuperscript{st} Salim Fares}
\IEEEauthorblockA{\textit{University of Passau} \\
Passau, Germany \\
salim.fares@uni-passau.de\\
}
\and
\IEEEauthorblockN{2\textsuperscript{nd} Steffen Herbold}
\IEEEauthorblockA{\textit{University of Passau} \\
Passau, Germany \\
steffen.herbold@uni-passau.de
}
}

\maketitle

\begin{abstract}

Software developers increasingly use AI agents that execute multi-step programming tasks on their own. However, it is unclear if and how this transfers to robotic engineering, where engineers program industrial robots in niche, vendor-specific languages and rely on visual feedback from simulation tools. Within this paper, we report a case study on the use of AI agents by an SME that programs industrial robots. Our results show that the agent did not make the engineers more productive, with the exception of a task that required the generation of highly repetitive code. However, the results with the agent improved over time, as the engineers learned how to use it and the agent was adapted to the domain. Moreover, the experience with AI mattered more for the success with the agent than the experience with robotic engineering. This shows that AI agents are not a plug-and-play solution. Instead, companies must train their engineers on the usage of agents, plan time for a learning curve, and ensure that agent configurations match the domain and developer requirements. The strong learning curve we observed further implies that studies on productivity gains with AI should be longitudinal, as they otherwise miss how engineers adopt agents in practice.
\end{abstract}


\section{Introduction}

Programming of industrial robots is a specialized sub-domain at the intersection of software development and industrial engineering: engineers write code in vendor-specific languages like ABB RAPID and check the movements of the robot visually in simulation tools like RobotStudio~\cite{ABB_RobotStudio}. Projects often differ from typical software development in multiple aspects, e.g., due to the use of visual programming and because solutions are typically adapted from previous programs rather than re-written from scratch. Thus, repetitive and error-prone tasks like renaming, shifting positions, or reversing paths are common~\cite{10.1145/3786583.3786869}. 

At the same time, software developers increasingly use AI agents that execute multi-step tasks on their own, instead of only completing code or answering questions in a chat~\cite{johnston2026shiftagenticaievidence}. However, the evidence on the productivity effects of such tools is inconclusive~\cite{becker2025measuringimpactearly2025ai, Sarkar2026AI, NBERw35275} and comes from mainstream software development. It is unclear if and how it transfers to robotic engineering. Work on Large Language Models (LLMs) for industrial automation evaluates the AI solutions in isolation, i.e., without engineers in the loop~\cite{Fakih_2024,10.1145/3786583.3786869}. To the best of our knowledge, there is no empirical evidence on whether agents help robotic engineers in industry with the range of their daily tasks, i.e., beyond isolated code changes. Similarly, we do not know how these engineers use agents and which problems they face when adopting them.

\begin{description}
    \item[\textbf{RQ1}] How does using agentic AI affect the completion, time, and quality of robotic engineering tasks compared to working without AI?
    \item[\textbf{RQ2}] How do engineers that are specialized in the programming of industrial robots use agentic AI and which problems do they face when adopting this new technology?
\end{description}

Our results indicate that the agent did not help the developers to solve tasks faster or with better functionality than without AI. The only exception is the generation of highly repetitive code. However, the solutions with the agent improved between the focus days, especially regarding the adherence to AKE's programming standards. We believe that this is due to both the learning curve of the developers and our re-design of the agent between the focus days, i.e., the configuration of the agent must also fit this setting. Regarding the usage, we found that the developers delegated all work to the agent, including trivial changes, used it like a chatbot instead of invoking the skills we prepared, and missed the visual feedback from the virtual station. Moreover, the experience with AI mattered more for the successful use of the agent than the experience with robotic engineering.

Based on these findings, we derive the following lessons for practitioners:
\begin{itemize}
    \item An on-premise agent with an open-weight model is technically feasible for SMEs with confidential data. However, it is not a plug-and-play solution that directly increases productivity.
    \item Experienced robotic engineers are not automatically experienced users of agents. Thus, SMEs should train the usage of agents, e.g., when to delegate and when to change code manually, and plan time for a learning curve.
    \item Repetitive and well-specified tasks, e.g., the generation of trajectories, are a better starting point than tasks that require a spatial understanding of the robot cell.
    \item Domain knowledge like reference manuals and programming standards should be available to the agent automatically, instead of relying on the developers to request it.
\end{itemize}

\section{Related Work}
Recent research has increasingly examined the effects of AI and LLMs workers' productivity including coding~\cite{cui2025the}, customer support~\cite{10.1093/qje/qjae044}, law~\cite{3ed03da3762842ed807d58637000d608} and many other fields. Most existing estimates focus on earlier-generation tools, such as autocomplete systems and chatbots, while developers increasingly adopt agentic tools~\cite{johnston2026shiftagenticaievidence}. Evidence on the productivity effects of these newer tools remains limited and inconclusive \cite{NBERw35275}. Becker et al.~\cite{becker2025measuringimpactearly2025ai} find that AI coding tools reduce the productivity of experienced developers, whereas other studies document substantial gains~\cite{Sarkar2026AI}. Chen et al.~\cite{chen2026artificial} show that AI tools increase coding activity within enterprises, although these increases translate only partially into greater output. Similarly, Demirer et al.~\cite{NBERw35275} demonstrate that successive generations of AI coding tools generate progressively larger productivity gains at the task level, with diminishing returns for more complex or higher-level tasks. 

In specialized industrial control and manufacturing environments, achieving verifiable code and reducing commissioning time are critical determinants of productivity. Fakih et al.~\cite{Fakih_2024} address this by introducing LLM4PLC, an agentic pipeline for industrial control programming that pairs generative models with external compilers, grammar checkers, and model verifiers. By offloading error detection and logic repair to an automated iterative agent loop, their approach increases compilation pass rates from 47\% to 72\% and drastically reduces engineering iteration cycles. However, this work was only conducted using the FischerTechnik Manufacturing TestBed, i.e., never validated in a real-life industrial setting.

In a complementary domain-specific application, Patel \cite{patel2024agentic} proposes an agentic AI framework for self-healing production lines with robotics, enabling autonomous fault diagnosis and correction in real time. His findings indicate a substantial decrease in downtime, with incidents often being resolved more than 80\% faster, and defects are almost entirely prevented from spreading to subsequent stages of production. This shows the potential of agentic systems for maintenance tasks, but does not look at the programming of new components. 

In our previous research~\cite{10.1145/3786583.3786869} we explored the capabilities of a general-purpose LLM in the industrial robotic manufacturing process through prompt engineering. We showed that an SME can utilize Llama 3.1 70B~\cite{grattafiori2024llama3herdmodels} to solve simple but recurring problems in ABB's RAPID through prompt engineering without the need for expensive solutions like training their own model. However, this work was limited to three controlled problems and evaluated on historic data.

To the best of our knowledge, there is no study on the general use of agentic AI  for robotic programming that spans multiple aspects, i.e., developing new components, modifying existing components to work in a different setting, documenting projects, and finding defects.

\section{Methodology}

We study the use of AI agents in a real industrial context, i.e., with real engineers, real tasks, and real constraints. This allows us to observe how agents perform in practice and how industrial engineers that are programming robotic software use them in their daily work. Methodologically, our work is therefore a case study in the sense of Wohlin et al.~\cite{wohlin2012experimentation}, i.e., we study a small sample of engineers and tasks in depth within their real working environment and we do not aim for statistical generalization. Instead, we want to collect rich quantitative and qualitative evidence on how the agent performs and how the engineers use it in this industrial context. 

\subsection{Industrial context}\label{sec:industrial-context}

We conducted our case study together with AKE Technologies, an SME that develops robotic automation solutions for the manufacturing processes of their customers. The robotic engineers at AKE program ABB robots in ABB RAPID. They develop and test their programs in virtual stations in RobotStudio~\cite{ABB_RobotStudio}, i.e., ABB's environment for simulation and offline programming, before they deploy them to the real robots at the customer.

Two constraints at AKE restrict how AI can be used. First, the code and data of customers are confidential and the use of cloud-based LLMs is not allowed. A local server with a single H100 GPU is available for inference. Second, AKE does not have the resources to develop or train a custom solution, i.e., any solution should be easy to maintain and rely on public models and agent harnesses. We believe that this setting is representative for many SMEs that consider adopting AI agents in confidential contexts with a limited budget.

The case study was conducted with the group of robotics programming at AKE. The \emph{team lead} and head of robotic programming served as our primary point of contact. He was also responsible to ensure that everything is embedded realistically in the development processes and defined the tasks for which we conducted direct measurements (Section~\ref{sec:use-cases}) and assessed the functional and non-functional quality of the solutions, according to AKE's standards (Section~\ref{sec:evaluation}). The four \emph{developers} working in the team solved tasks with and without the use of AI (Section~\ref{sec:focus-days}). 

Our role as researchers was to provide and monitor the AI solution and define the scientific instruments of the case study, i.e., criteria to assess the solutions, the questionnaires, and the workflow of the focus days. We note that we did not have any influence on the tasks or their assessment, which was done by the team lead according to AKE's standards.

\subsection{Tasks}\label{sec:use-cases}

For this case study, we extended the use cases from our prior work with AKE~\cite{10.1145/3786583.3786869} such that they cover the range of tasks that engineers encounter in their daily work, i.e., refactoring and extension of existing programs, documentation, debugging, path inversion, trajectory generation, and 3D stacking (Sections~\ref{sec:task1}--\ref{sec:task6}). For each of these activities, the team lead defined concrete tasks from actual project contexts as study objects for us (Section~\ref{sec:industrial-context}). The developers were given natural language descriptions of the tasks that followed AKE's style guides and conventions in exactly the same manner as real projects, i.e., while we focused on specific tasks, the actual tasks occur as is in the daily development. The task descriptions were in German, which is the mother tongue of all developers. All tasks are non-trivial activities, for which the estimated effort is between 30 and 60 minutes.

Overall, we defined six tasks. Tasks~1, 4, and~6 are related to the use cases from our prior work~\cite{10.1145/3786583.3786869}, i.e., the renaming of arguments as key activity of refactoring projects (Task~1), the reversal of movements (Task~4), and working with offsets of positions (Task~6). For these tasks, we already know that a general-purpose LLM can perform them through prompt engineering, and these tasks allow us to generalize the findings from individual changes to larger operations in real-life contexts guided by a human developer. The other three tasks were not part of our prior work, i.e., the generation of documentation for code (Task~2), the identification of mistakes (Task~3), and the creation of new movement trajectories (Task~5). In general, the new tasks require similar skills as the existing tasks. However, they focus more on the creation of new information, i.e., documents, finding unknown errors, and new trajectories, than on the modification of existing code.

\subsubsection{Task 1: Refactoring and extension}\label{sec:task1}
The engineers extend an existing program with new code that is almost identical to code that already exists, i.e., they copy existing routines, apply small changes, e.g., the renaming of variables, and call the new routines at the correct positions in the program. This is one of the most common tasks at AKE: most projects build on code from previously finished projects due to the similarity between production lines. While each individual change is simple, the challenge is to make all changes consistently and to integrate the new code correctly into the existing program flow.

\subsubsection{Task 2: Documentation}\label{sec:task2}
The engineers write external textual documentation of a robotic program that follows AKE's standards. Concretely, they fill a template provided by AKE that requires system metadata (e.g., the project name, the robot type), all defined coordinate systems and robotic tools, a description of the workflow of the program, and a list of all movement routines with a short description of their purpose. AKE needs such documentation when customers request an explanation of the implemented solution. This is the only task that does not require the engineers to write RAPID code. Instead, they must understand the existing code and the workflow and describe both in natural language.

\subsubsection{Task 3: Debugging}\label{sec:task3}
Debugging is a very common task in any development process. The engineers must find and fix eight subtle errors in a single code module, i.e., two syntax errors and six semantic errors. The two types of errors require different knowledge. The syntax errors require familiarity with ABB RAPID, while the semantic errors require that the engineers understand the workflow of the code and the behavior of the robot. The errors were induced by the team lead based on common error patterns that the engineers encounter.

\subsubsection{Task 4: Path inversion}\label{sec:task4}
The engineers are given a forward sequence of three movement routines and must implement the inverse sequence. Switching the order of the movements is not sufficient for this. Depending on the path, the engineers may also have to change the type of some movements, add new movements, or remove existing movements. Thus, the task requires not only code changes, but also an understanding of how the robot moves.

\subsubsection{Task 5: Trajectory matrix generation}\label{sec:task5}
The engineers are only given the coordinates of multiple positions. They must create the complete movement paths between these positions, i.e., two movement routines (forward and backward) between each pair of positions from scratch, without without existing code to start from. This is common for situations where a completely new component is introduced into a production line. While each individual routine is relatively simple, the number of routines grows quickly with the number of positions. Thus, the task is repetitive and prone to small mistakes.

\subsubsection{Task 6: 3D stacking}\label{sec:task6}
The engineers must program the robot to stack batteries on top of each other. The movements are repetitive, but with different offsets for the pick-up and place-down positions of each battery. The team lead reported this task as the most difficult of our tasks, because it requires a mathematical computation of the movements' offset and planning an efficient movement order to stack the batteries.

\subsection{Customized AI Agent}\label{sec:agent-config}

Our goal was to provide the developers with an agent that is ready to use, i.e., the developers should not have to set up or configure the agent themselves. Due to the constraints at AKE (Section~\ref{sec:industrial-context}), we restricted ourselves to an off-the-shelf agent and an open-weight model that run fully on-premise on AKE's existing hardware. 

We use Claude Code~\cite{Anthropic2026ClaudeCode} in the terminal as agent, because it provides the agentic loop out of the box, i.e., it can read files, write code, execute shell commands, and check its own changes without the need for custom agent scaffolding. Moreover, Claude Code can be connected to any server with an Anthropic-compatible API. As an LLM, we use GPT-OSS 120B~\cite{openai2025gptoss120bgptoss20bmodel}, which we serve through Ollama~\cite{Ollama2023}. 

Before the focus days, we validated that this setup is capable of solving the tasks in general, i.e., that it generates syntactically correct RAPID code and that it understands the tasks. We did not require that the agent solves the tasks fully autonomously. Instead, it was sufficient that the agent generated reasonable outputs that were on a trajectory towards a correct solution. During this testing period, we also considered other open-weight models, including Qwen3.6 35B~\cite{qwen36_35b_a3b}, Gemma4 31B~\cite{gemmateam2026gemma4technicalreport}, GLM-4.7-Flash Q4~\cite{glm47flashq4km} and Laguna XS.2 Q4~\cite{laguna_xs2_q4km}, but we found that GPT-OSS 120B performed best for our tasks. We note that we did not perform a systematic evaluation of different models, because this was not the focus of our case study. Instead, we focused on the interaction between the developers and the agent, which is independent of the specific model.

For the first focus day, we provided three skills, which we derived from the use cases of our prior work~\cite{10.1145/3786583.3786869}: \emph{argument-modification} renames variables in a movement routine (related to Task~1), \emph{offset} adds an offset instruction to a movement routine (related to Task~6), and \emph{reverse} reverses a movement routine (related to Task~4). The developers had to select and invoke the skills themselves whenever they deemed them useful. We note that the tasks were designed such that these skills are frequently required.

However, we observed that the developers did not use the skills as intended on the first focus day (Section~\ref{sec:results-day1}). Therefore, we re-designed the configuration of the agent for the second focus day such that the developers no longer had to select skills themselves. Instead, we configured a local project context that Claude uses automatically. This context consists of the ABB RAPID reference manual as a markdown file, a PDF document of the task descriptions (the same as the one that the developers received), and an initialized CLAUDE.md file that contains the structure of the project and the rules for generating RAPID code according to the manual. This way, the agent has access to the domain knowledge without the developers having to request it explicitly.

\subsection{Focus days}\label{sec:focus-days}

The main part of our case study are two focus days, during which we observed how the developers solved the tasks with and without the agent. On each focus day, AKE gathered the engineers of the robotic programming team, who reserved the full day for working on the tasks. The developers worked on their own machines with their usual tools, e.g., RobotStudio, and their usual development setup. However, instead of working on their usual projects, they solved the tasks we provided. Thus, they more or less had a regular work day with mixed tasks, as can happen whenever there is demand from multiple projects at the same time. This setting gives us some control over what we observe, while the engineers still work in their usual environment.

We split the developers into two groups. On the first focus day, Group~A worked without AI in the morning and with the agent in the afternoon, while Group~B worked with the agent in the morning and without AI in the afternoon. On the second focus day, the groups swapped. This is essentially an AB/BA crossover design, which allows us to reduce the influence of individual differences between the developers. We note, however, that the small number of developers does not allow for statistical generalization.

Since each developer works on every type of task on both focus days, i.e., once with the agent and once without AI, we did not use the exact same tasks twice. Instead, the tasks of the second focus day are of the same type as on the first focus day, but with different descriptions (e.g., on the first focus day for Task~4 the forward movement of a path was given and the backward movement path had to be implemented to go through, on the second focus day the positions were changed and renamed, and the backward movement path was given and the forward had to be implemented for a drilling sequence). Thus, the type and the difficulty of the tasks remain the same, while the developers cannot simply reproduce their solutions from the first focus day.

Each focus day started with a short introduction to the agent by the first author, which covered a general description of the agent. On the first focus day, the introduction additionally explained which skills we prepared for the developers and how to use them. On the second focus day, the introduction instead explained how we configured the local project context with the reference manual and the task description, which removes the need for the developers to select skills themselves. Afterwards, the developers worked on the tasks, three in the morning and three in the afternoon, with a lunch break in between. For each developer and task, we collected the time, the answers to the task questionnaire, a backup of the resulting RobotStudio project, and, if the developer used the agent, the chat history with the agent (Sections~\ref{sec:questionnaires} and~\ref{sec:evaluation}). The tasks were time-boxed, i.e., if developers did not finish a task within the time limit, they had to stop and move on to the next task. The time limits were set by the team lead based on the task complexity. 

Between the two focus days, there was a break of eight weeks. During this period, we encouraged the developers to use the agent within their daily work however they wanted. 

\subsection{Questionnaires}\label{sec:questionnaires}

As part of our case study, we wanted to understand how the developers perceive the agent, e.g., which problems they encounter and whether their expectations are met. Thus, we defined questionnaires that the developers answered at different points of the focus days (Section~\ref{sec:focus-days}). The questionnaires were written in German and handed out on paper. Unless stated otherwise, the closed questions use a five-point Likert scale from 1 (strongly disagree) to 5 (strongly agree).

The developers filled the \emph{pre-questionnaire} once at the beginning of the first focus day. It contains demographic questions, questions about the experience of the developers with AI (self-rated knowledge and frequency of use), and questions about their expectations of the agent. We measure the expectations as agreement with five statements, i.e., that using AI 1) leads to better results, 2) adheres to AKE's programming standards, 3) speeds up the development, and 4) understands the tasks correctly, as well as 5) that the developer wants to use AI in their work. Additionally, the developers rated whether they expect the agent to be easy to use and the tasks to be difficult with and without AI.

The developers filled the \emph{task questionnaire} after each task, i.e., six times per focus day. It asks for the perceived difficulty and realism of the task, the self-reported time spent on thinking about the solution (brainstorming) and on the actual implementation, and the problems the developer faced during the task. On the second focus day, the developers had solved each type of task once with and once without the agent. Therefore, the task questionnaire of the second focus day additionally asks for the agreement that the agent avoided problems of the development without AI and that it led to problems that the development without AI avoided, each with a free-text field for examples.

The developers filled the \emph{post-questionnaire} at the end of each focus day. It repeats the five statements about the expectations from the pre-questionnaire, this time about the actual experience with the agent, and asks whether the agent was helpful for solving the tasks. Two free-text questions ask what the agent was helpful for and which problems occurred. This way, we can compare the expectations of the developers with their experience after each focus day.

\subsection{Evaluation of task solutions}\label{sec:evaluation}

We evaluate the solutions of the developers with respect to the time they required and the quality of their results. For the quality, we combine a manual assessment by the team lead with an automated validation of the RAPID code. Additionally, we use the questionnaires to understand how the developers perceived and used the agent.

The first author recorded the time in minutes for each developer and task, i.e., from the start of the task until the developer reported that they were finished. The time limits for the tasks were 60 minutes for Task~1, 30 minutes for Tasks~2--4, and 45 minutes for Tasks~5 and~6. We report whether the developers finished the task in time.

After each focus day, the team lead assessed the resulting RobotStudio backups of all developers and tasks. We defined three criteria for this together with the team lead. 
\begin{itemize}
    \item The \emph{functionality}, i.e., whether the solution fulfills the requirements of the task and the robot behaves correctly in the virtual station. 
    \item The \emph{code quality}, i.e., whether the code adheres to AKE's programming standards.
    \item The \emph{readability}, i.e., whether other engineers at AKE can easily understand the code, e.g., through naming and comments. 
\end{itemize}
The team lead rated each criterion on a scale from \emph{1} (Criterion not met) to \emph{5} (Criterion exceeds expectations significantly) relative to the expectations of AKE for a solution of an experienced engineer. We note that the assessment was not blind, i.e., the team lead knew which solutions were created with the agent and which without AI. 
For the debugging task (Task~3), these criteria are not meaningful, because the developers modify existing code instead of creating new code. Instead, the team lead counted how many of the eight injected errors the developers fixed and how many remained, separately for the two syntax errors and the six semantic errors.

Finally, we used the custom validator from our prior work~\cite{10.1145/3786583.3786869} to automatically check the RAPID code of all solutions for syntactic violations of AKE's programming standards, e.g., naming conventions.

\section{Results}\label{sec:results}

Table~\ref{tab:results_bold} summarizes the results for all tasks. We refer to the developers as D1--D4. We note that D3 could not participate in the second focus day for personal reasons unrelated to the case study. Thus, we have 24 task solutions from the first focus day and 18 from the second focus day. Regardless of the focus day and the use of the agent, the developers rated the tasks as realistic, i.e., the mean agreement over all task questionnaires is 4.55 and no developer disagreed for any task. Thus, the tasks reflect the daily work of the developers.

\begin{table*}[htbp]
\caption{Time and assessment by the team lead per task, developer, and focus day. The time is shown in minutes. Values written in bold mean that AI was used for this task on this focus day. The assessment uses 1 (criterion not met), 2 (partially met), 3 (meets expectations), 4 (partially exceeds expectations), 5 (criterion exceeds expectations significantly). -- means that a value is missing because it is not evaluated for a task or because D3 was not available for the second focus day.}
\label{tab:results_bold}
\centering
\begin{tabular}{|c|c|cc|cc|cc|cc|cc|}
    \hline
    \multirow{2}{*}{Task} & \multirow{2}{*}{Developer}
    & \multicolumn{2}{|c|}{Time}
    & \multicolumn{2}{|c|}{Functionality}
    & \multicolumn{2}{|c|}{Code Quality}
    & \multicolumn{2}{|c|}{Readability}
    & \multicolumn{2}{|c|}{\#Prompts}\\
    \cline{3-12}
    & & FD1 & FD2 & FD1 & FD2 & FD1 & FD2 & FD1 & FD2 & FD1 & FD2 \\
    \hline
    \multirow{4}{*}{Task 1}
    
    & D1 & 38 & \textbf{60} & 3 & \textbf{1} & 3 & \textbf{1} & 3 & \textbf{2} & -- & \textbf{11}\\
    \cline{2-12}
    & D2 & \textbf{60} & 18 & \textbf{1} & 3 & \textbf{1} & 3 & \textbf{1} & 3 & \textbf{12} & -- \\
    \cline{2-12}
    & D3 & 42 & -- & 3 & -- & 3 & -- & 4 & -- & -- & -- \\
    \cline{2-12}
    & D4 & \textbf{60} & 33 & \textbf{1} & 3 & \textbf{1} & 3 & \textbf{1} & 3 & \textbf{7} & --\\
    \hline

    \multirow{4}{*}{Task 2}
    & D1 & 30 & \textbf{30} & 1 & \textbf{2} & 2 & \textbf{2} & 3 & \textbf{2} & -- & \textbf{5}\\
    \cline{2-12}
    & D2 & \textbf{30} & 30 & \textbf{2} & 2 & \textbf{2} & 3 & \textbf{3} & 3 & \textbf{2} & --\\
    \cline{2-12}
    & D3 & 30 & -- & 1 & -- & 2 & -- & 2 & -- & -- & -- \\
    \cline{2-12}
    & D4 & \textbf{30} & 30 & \textbf{2} & 2 & \textbf{1} & 2 & \textbf{3} & 3 &\textbf{4} & --\\
    \hline

    \multirow{4}{*}{Task 3}
    & D1 & 24 & \textbf{30} & -- & -- & -- & -- & -- & -- & -- & \textbf{8}\\
    \cline{2-12}
    & D2 & \textbf{27} & 13 & -- & -- & -- & -- & -- & --  & \textbf{5} & --\\
    \cline{2-12}
    & D3 & 26 & -- & -- & -- & -- & -- & -- & -- & -- & --\\
    \cline{2-12}
    & D4 & \textbf{30} & 17 & -- & -- & -- & -- & -- & -- & \textbf{5} & --\\
    \hline

    \multirow{4}{*}{Task 4}
    & D1 & \textbf{30} & 30 & \textbf{1} & 2 & \textbf{1} & 2 & \textbf{2} & 3 & \textbf{11} & --\\
    \cline{2-12}
    & D2 & 14 & \textbf{30} & 3 & \textbf{2} & 3 & \textbf{2} & 3 & \textbf{3} & -- & \textbf{7}\\
    \cline{2-12}
    & D3 & \textbf{30} & -- & \textbf{1} & -- & \textbf{2} & -- & \textbf{3} & -- & \textbf{8} & --\\
    \cline{2-12}
    & D4 & 26 & \textbf{30} & 3 & \textbf{1} & 3 & \textbf{2} & 3 & \textbf{3} & -- & \textbf{11}\\
    \hline

    \multirow{4}{*}{Task 5}
    & D1 & \textbf{20} & 15 & \textbf{3} & 2 & \textbf{3} & 3 & \textbf{2} & 2 & \textbf{13} & --\\
    \cline{2-12}
    & D2 & 18 & \textbf{15} & 3 & \textbf{3} & 3 & \textbf{3} & 3 & \textbf{3} & -- & \textbf{8}\\
    \cline{2-12}
    & D3 & \textbf{38} & -- & \textbf{3} & -- & \textbf{3} & -- & \textbf{2} & -- & \textbf{16} & --\\
    \cline{2-12}
    & D4 & 23 & \textbf{45} & 3 & \textbf{3} & 3 & \textbf{3} & 3 & \textbf{2} & -- & \textbf{11}\\
    \hline

    \multirow{4}{*}{Task 6}
    & D1 & \textbf{45} & 45 & \textbf{2} & 2 & \textbf{2} & 2 & \textbf{1} & 2 & \textbf{7} & --\\
    \cline{2-12}
    & D2 & 44 & \textbf{45} & 3 & \textbf{1} & 2 & \textbf{2} & 3 & \textbf{2} & -- & \textbf{13}\\
    \cline{2-12}
    & D3 & \textbf{45} & -- & \textbf{1} & -- & \textbf{2} & -- & \textbf{2} & -- & \textbf{18} & --\\
    \cline{2-12}
    & D4 & 45 & \textbf{45} & 1 & \textbf{1} & 1 & \textbf{1} & 2 & \textbf{2} & -- & \textbf{9}\\
    \hline

\end{tabular}
\end{table*}

\subsection{Participants and expectations}\label{sec:results-participants}

All four developers are experienced robotic engineers with 7 to 14 years of experience with RobotStudio. D1 has the most experience (14 years), followed by D4 (10 years with RobotStudio, 13 years at AKE), D2 (7.5 years), and D3 (7 years). In contrast, the experience with AI is more diverse and does not follow the robotics experience. D2 rates their AI knowledge as good and uses AI almost daily. The other developers have basic knowledge about AI, but differ in how often they use it: D1 uses AI weekly, D3 a few times per month, and D4 rarely or never. Thus, D2 is the developer with the most AI experience, while D1 and D4 are the most experienced robotic engineers.

\begin{figure}[!t]
\centerline{\includegraphics[width=0.45\textwidth]{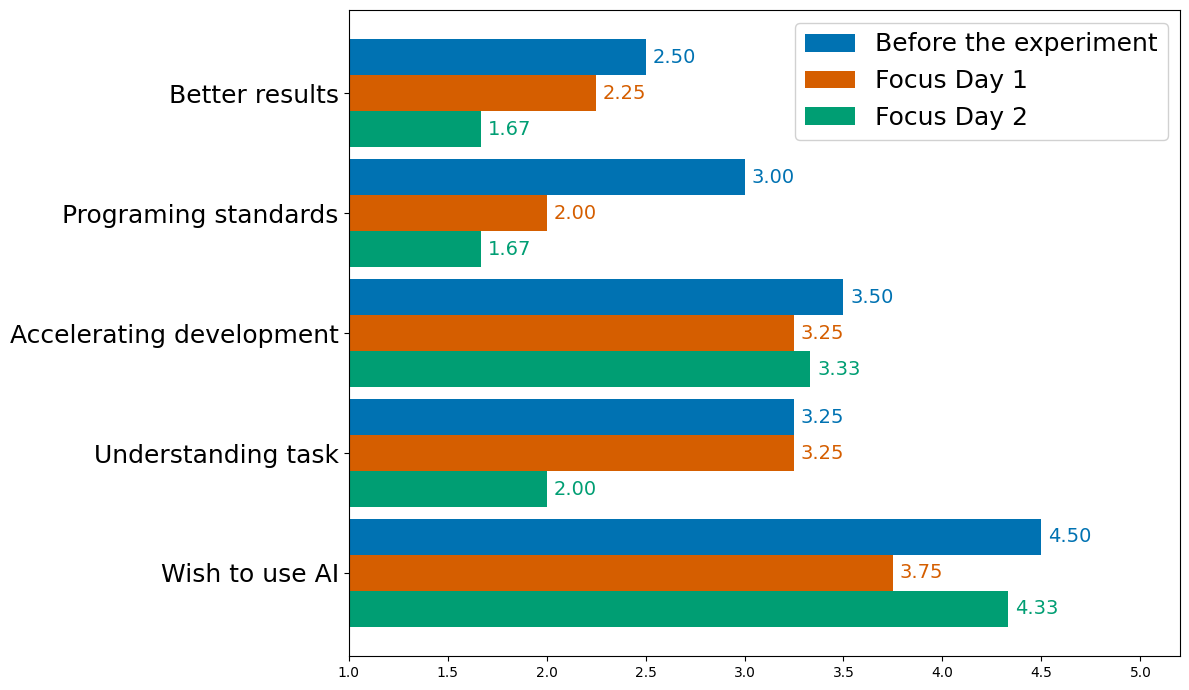}}
\caption{Mean agreement of the developers with the five statements before the first focus day and after each focus day (1 = strongly disagree, 5 = strongly agree). D3 did not answer after the second focus day.}
\label{fig:expectation_vs_reality}
\end{figure}

Before the first focus day, the developers were open towards the agent (Figure~\ref{fig:expectation_vs_reality}). They wanted to use AI agents for robotic projects (mean 4.5) and rather expected that the agent speeds up the development (3.5) and understands the tasks correctly (3.25). However, they were neutral regarding the adherence to AKE's programming standards (3.0) and whether their results become better (2.5). D1 and D2 expected that the agent is easy to use (5) and that the tasks are not difficult without AI (1), while D3 and D4 were neutral on both statements. Nobody expected that the tasks are difficult with the agent (mean 2.75). Overall, the developers started with moderately positive expectations regarding the speed of the development, but not regarding the quality of their results.

\begin{figure}[!t]
\centerline{\includegraphics[width=0.45\textwidth]{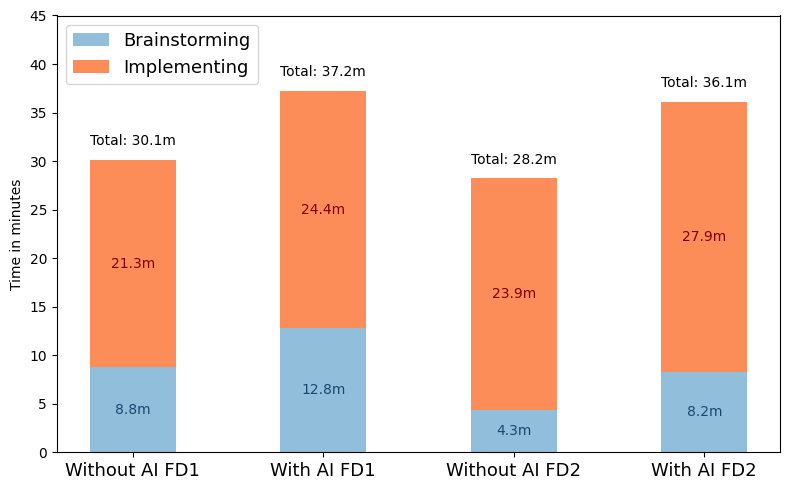}}
\caption{Mean self-reported time for brainstorming and implementation per task with the agent (With AI) and without the agent (Without AI) on both focus days.}
\label{fig:brainstorming_working_time}
\end{figure}

\subsection{First focus day}\label{sec:results-day1}

\paragraph{Completion and time}
Table~\ref{tab:results_bold} shows that the agent did not help the developers to solve the tasks faster on the first focus day. With the agent, the developers finished only 3 of 12 tasks within the time limit, i.e., D2 the debugging task (Task~3) and D1 and D3 the trajectory matrix generation (Task~5). Without AI, they finished 9 of 12 tasks. Moreover, the developers with the agent required at least as much time as the developers without AI for every task. The difference is largest for Tasks~1 and~4: both developers with the agent hit the time limit, while the developers without AI finished Task~1 after 38 and 42 of 60 minutes and Task~4 after 14 and 26 of 30 minutes. We note that some tasks were also hard without AI, i.e., nobody finished the documentation task (Task~2) in time and only D2 finished the 3D stacking task (Task~6). Task~5 is the only task that all developers finished.

\paragraph{Quality of solutions}
The assessment by the team lead is similar. For Tasks~1, 4, and~6, the team lead rated the solutions without AI better than those with the agent in all three criteria. For example, both solutions of Task~1 with the agent did not meet any criterion, while both solutions without AI met the expectations in all criteria. On average over all tasks, the functionality is 1.7 with the agent and 2.3 without AI, the code quality 1.8 and 2.5, and the readability 2.1 and 2.8. There are only two exceptions. For the documentation task (Task~2), both solutions with the agent partially met the functionality criterion, while both solutions without AI did not meet it. For Task~5, the solutions with the agent met the expectations for functionality and code quality, i.e., the same as without AI, but only partially met the readability criterion.

The validator confirms that the solutions with the agent adhere less to AKE's programming standards. The validator found 25 violations in the solutions with the agent and 11 in the solutions without AI. Only 2 of the 10 checked solutions with the agent are free of violations, in comparison to 6 of 10 without AI. The agent also did not help with debugging. In Task~3, no developer found all six logic errors. With the agent, 7 logic errors and 2 syntax errors remained, without AI 5 logic errors and no syntax errors. The two syntax errors remained in the solution of D4, who did not finish the task. Thus, the solutions with the agent were worse than the solutions without AI for four of the six tasks.

\paragraph{Developers' perception}
The perception of the developers matches these results. The developers perceived the tasks as slightly more difficult with the agent (mean 2.75) than without AI (2.5). However, this differs between the tasks: with the agent, the developers found Tasks~1 and~3 easier, but Tasks~2 and~4--6 harder than without AI. With the agent, the developers also reported more time for brainstorming (12.8 instead of 8.8 minutes) and for the implementation (24.4 instead of 21.3 minutes) (Figure~\ref{fig:brainstorming_working_time}). Without AI, the developers mostly reported problems that are typical for their daily work, e.g., adapting code after copy-and-paste (3 mentions), searching for information, and finding errors. With the agent, the most frequent problem was errors of the agent (4 mentions), followed by using the agent and checking its results (2 mentions each). This indicates that the agent replaced the usual problems with new ones, instead of removing them.

After the first focus day, the agreement with all statements except the correct understanding of the tasks dropped below the expectations (Figure~\ref{fig:expectation_vs_reality}), most notably for the programming standards (from 3.0 to 2.0) and the desire to use AI (from 4.5 to 3.75). The developers were neutral regarding the helpfulness of the agent for solving the tasks (mean 3.0). In the free-text answers, the developers found the agent helpful for typing, documentation, and finding logic errors. As problems, they named that the agent is not user-friendly and new to them (2 mentions each), that it does not know the ABB documentation, and that it has no access to RobotStudio. The usability problems mainly refer to the terminal interface of Claude Code: the developers usually work with graphical tools like RobotStudio and reported that working in the terminal was new and unfriendly for them. Thus, the first focus day lowered the expectations of the developers, but did not lead to a rejection of the agent.

\paragraph{Use of the agent}\label{sec:results-day1-usage}
The chat histories and our observations during the focus day explain, at least partially, why the agent did not help. The developers did not invoke any of the three skills we prepared (with one single exception from D3 on Task~4), and they did not provide the agent with reference code from the project to explain the task or the given code. Instead, the developers used the agent for every change, even for trivial ones like adding or removing a comma. As a consequence, the agent regenerated large parts of the code, sometimes whole modules, again and again. The conversations became very long, and towards their end, the agent started to hallucinate and drifted away from correct intermediate solutions. We observed that the first answers of the agent were often a good starting point. However, the repeated regeneration for small changes negated the time saved through these first answers. Table \ref{tab:results_bold} shows the number of prompts per task for each developer. We can see that the developers needed fewer prompts for Task~2 and 3 where the developers asked the agent to scan and fill information or fix the code, whereas they interacted more with the agent for the tasks that required multiple actions and adjustments, like Task~6.

Moreover, we note that the developers who used AI at least occasionally before the case study (D1, D2, D3) finished one task each with the agent, while D4, who rarely or never uses AI, finished none.

\subsection{Second focus day}\label{sec:results-day2}

\paragraph{Completion and time}
On the second focus day, the completion rate with the agent did not improve (Table~\ref{tab:results_bold}). With the agent, the developers finished only 1 of 9 tasks within the time limit, i.e., D2 the trajectory matrix generation (Task~5). Without AI, they finished 5 of 9 tasks. The solution of D2 for Task~5 took 15 minutes, which is less than the 18 minutes D2 required for the same type of task without AI on the first focus day. This is also as fast as D1, the most experienced robotic engineer, without AI on the second focus day. Otherwise, the developers with the agent again required at least as much time as the developers without AI. For example, D1 hit the time limit of Task~1 with the agent, while D2 and D4 finished the same task without AI after 18 and 33 minutes. Nobody finished the documentation task (Task~2) or the 3D stacking task (Task~6).

\paragraph{Quality of solutions}
The assessment by the team lead is more mixed than on the first focus day. For Task~5, both solutions with the agent met the expectations for functionality and code quality, while the solution without AI only partially met the functionality criterion. For Task~2, the solutions with and without the agent partially met the functionality criterion. However, for Tasks~1, 4, and~6, the functionality of all solutions with the agent did not meet the criterion, while the solutions without AI met the expectations (Task~1) or partially met them (Tasks~4 and~6). On average over all tasks, the functionality with the agent is 1.8 (without AI 2.3), the code quality 2.0 (2.3), and the readability 2.4 (2.7). Thus, the gap to the solutions without AI became smaller for code quality and readability, while the gap for functionality mostly remains.

The validator shows a clearer change. It found 16 violations in the solutions with the agent and 15 in the solutions without AI. In contrast to the first focus day, 5 of 8 checked solutions with the agent are free of violations, but only 1 of 7 without AI. We note that the violations without AI are mostly wrong IDs in Tasks~1 and~3 (6 violations) and the solution of D1 for Task~4 (8 violations). Moreover, six wrong movements in Task~4 occur in all three solutions, with and without the agent. In the debugging task (Task~3), both developers without AI fixed all errors, while D1 with the agent fixed the syntax errors, but six logic errors remained.

\paragraph{Developers' perception}
With the agent, the developers perceived the tasks as more difficult (mean 3.2) than without AI (2.1). Only the path inversion (Task~4) was perceived as easier with the agent (1.5 instead of 3.0). The developers reported less brainstorming time than on the first focus day, both with the agent (8.2 minutes) and without AI (4.3 minutes) (Figure~\ref{fig:brainstorming_working_time}). With the agent, the most frequent problem was to explain the task to the agent (4 mentions). Errors of the agent and problems with using the agent, which dominated the first focus day, were mentioned rarely. Without AI, the developers reported the same problems as during the first focus day. Thus, the problems with the agent changed, while the problems without AI remained the same. 

Since the developers solved each type of task with the other condition on the first focus day, they also rated whether the agent avoided problems they had without AI and whether the agent led to problems that manual work avoided. On average, the developers were neutral (means 3.3 and 3.4 over 15 answers), but the answers range from 1 to 5. We note that D1 only answered after the tasks without AI. The agreement that the agent avoided problems is highest for Task~5 (5, 5, and 4), which matches the results with respect to the functionality.

As examples for avoided problems, the developers named faster results, fewer errors during monotonous typing, more professional wording of the documentation, and checking the numbering of positions. As new problems, they named hallucinations, wrong renamings, syntax errors, the need to write exact prompts, that sorting routines was impossible, and that the agent only sees the code, but not the real conditions of the robot cell. Thus, the developers see both benefits and drawbacks of the agent, which depend on the task.

After the second focus day, the agreement that the agent understands the tasks correctly (2.0), and the agreement with better results (1.67) and adherence to the programming standards (1.67) all dropped (Figure~\ref{fig:expectation_vs_reality}). In contrast, the desire to use AI (4.33) and the expected speed-up (3.33) increased slightly in comparison to the first focus day. However, this comparison is affected by the absence of D3, who had the lowest agreement after the first focus day. For D1, D2, and D4 only, the desire to use AI increased from 4.0 to 4.33 and the agreement that the agent was helpful from 3.0 to 3.33, while the expected speed-up decreased from 3.67 to 3.33. In the free-text answers, the developers described the agent as helpful for searching information, documentation, and rewriting routines. They named server problems and explaining the task to the agent as problems. Nobody named the usability of the agent or the missing ABB documentation as a problem anymore. Thus, the developers became more critical regarding the quality of the results, while their desire to use the agent remained high.

\paragraph{Use of the agent}\label{sec:results-day2-usage}
We observed that the developers were more interested in the agent than on the first focus day and asked about commands and skills. We can see for Tasks~4 and~5 where the reversing skill is automatically activated, they required fewer prompts than on first focus day as shown in Table~\ref{tab:results_bold}. However, the chat histories show that the developers still used the agent for every change, including trivial ones, which again led to long conversations in which the agent drifted away from correct intermediate solutions. Thus, the re-design of the agent addressed the missing domain knowledge, but not this usage pattern.

\subsection{Comparison of the focus days}\label{sec:results-comparison}

Overall, the comparison of the two focus days shows only a partial improvement for the agent. The completion rate with the agent did not improve, i.e., 3 of 12 tasks on the first and 1 of 9 tasks on the second focus day, while the completion rate without AI was similar on both focus days (9 of 12 and 5 of 9). The functionality of the solutions with the agent did not improve. However, the code quality (from 0.8 to 1.0), the readability (from 1.0 to 1.4), and the adherence to the programming standards according to the validator improved, i.e., 16 instead of 25 violations and 5 of 8 instead of 2 of 10 solutions without violations. We believe that the better adherence to the programming standards due to the reference manual in the project context of the re-designed agent.

The trajectory matrix generation (Task~5) is the only task for which the agent led to the same or better results than without AI on both focus days. On the second focus day, D2 even finished this task faster with the agent than without AI on the first focus day. Moreover, the problems with the agent shifted from errors of the agent and the usage of the agent to explaining the task to the agent, while the developers reported the same problems without AI on both focus days. However, the perceived difficulty with the agent did not decrease: the developers perceived four of six tasks as harder with the agent on the second focus day than on the first focus day, and only the path inversion (Task~4) as easier. 

Regarding the experience of the developers, the only task completed with the agent on the second focus day was completed by D2, the developer with the most AI experience. In contrast, D1, the most experienced robotic engineer, did not finish any task with the agent on the second focus day, and D4, who rarely or never uses AI, did not finish any task with the agent on either focus day. Without AI, all developers finished tasks on both focus days. The post-questionnaires show a similar pattern. While the mean agreement with the statements decreased in comparison to the expectations, the individual developers differ. D4 is the only developer whose agreement decreased for all five statements after the first focus day and did not recover after the second focus day. In contrast, D1 and D2 each rated at least one statement higher after the first focus day than before, e.g., D1 agreed more that their results are better with the agent (from 2 to 3) and D2 agreed more that the agent speeds up the development (from 4 to 5). Thus, the developer with the least AI experience was also the one who was disappointed by the agent the most. This indicates that the experience with AI matters at least as much as the robotics experience for the successful use of the agent.

\section{Discussion}\label{sec:discussion}

\subsection{RQ1: Completion, time, and quality}\label{sec:disc-rq1}

Overall, our results show that the agent did not help the developers to complete more tasks, to complete them faster, or to create solutions with better functionality. This is not what we expected, because our prior work showed that an LLM can perform key steps of these tasks~\cite{10.1145/3786583.3786869} and related literature also indicated positive findings in other settings~\cite{Fakih_2024, patel2024agentic}. The likely cause is the difference between the clean, benchmark-like setting of our prior work and the real-world setting with developers who have different expectations and levels of experience with AI agents.

Notably, the agent performed worst where we expected it to perform best: for Tasks~1, 4, and~6, i.e., the tasks that are closest to our prior work. We see two potential interpretations. First, the capability of an LLM to perform small, isolated changes does not transfer to larger tasks in realistic settings, in which many such changes must be planned and combined within an existing project. Second, even if the agent can be used for such tasks, this does not mean that it is more efficient than experienced developers. However, neither interpretation implies that the agent cannot perform better in these settings. Our results also show that the inexperience of the developers with agents was a major factor (Section~\ref{sec:disc-rq2}) and that the gap to the solutions without AI became smaller for code quality and readability between the focus days, even though the gap for functionality mostly remained.

Consequently, we expect that the productivity with agents is not fixed, but shifts over time with training and experience. Our two focus days only provide two snapshots of this development. Thus, longitudinal studies are needed to understand how the productivity evolves and how developers can best be supported during this transition.

Still, we observed positive aspects of the agent. On the second focus day, the solutions with the agent adhered better to AKE's programming standards, i.e., 5 of 8 solutions were free of violations in comparison to 2 of 10 on the first focus day. Since this is (at least partially) also caused by our re-configuration of the agent, this shows that the learning curve does not only concern the developers. Thus, context-specific configurations are an important factor that must be carefully considered and controlled when assessing the performance of agents in real-world settings.

Moreover, the developers appreciated on both focus days that the agent reduced repetitive work, e.g., monotonous typing. This is also reflected by the trajectory matrix generation (Task~5), the only task for which the agent led to the same or better results than working without AI on both focus days. Once the task is well understood and the trajectories are known to the developers, the agent effectively supports the implementation of highly repetitive code. This indicates that the agent is most useful when the developers can precisely specify what they want and the remaining work is mostly the writing of code.

\begin{rqanswer}{Answer to RQ1}
The agent did not improve productivity, except for a highly repetitive task. However, performance with the agent improved between the focus days, which indicates that the configuration of the agent and the experience of the developers affect the outcomes. This indicates that better performance with more experience can still be expected.
\end{rqanswer}

\subsection{RQ2: Use of the agent and adoption problems}\label{sec:disc-rq2}

Our results indicate that the main problem was not only the capability of the model, but also how the developers used the agent. Experienced robotic engineers are not automatically experienced users of agents. Moreover, robotic development relies on visual tools and differs from the mostly text-based programming in other areas of software engineering, for which agents like Claude Code are designed.

First, the developers overused the agent. Instead of using it as an assistant, they delegated all work to it, including trivial changes such as adding a comma. This is not only inefficient, but also leads to long conversations, in which the agent drifted away from correct intermediate solutions. This is in line with the known degradation of LLMs for long contexts~\cite{liu-etal-2024-lost}. We observed this usage pattern on both focus days, i.e., the re-design of the agent did not address it.

Second, the developers were used to chat-based AI systems and used the agent in the same manner, even though the introduction explained how to invoke the skills. As a consequence, the developers ignored the skills we prepared, with a single exception, which contributed to the problems on the first focus day. This shows that the design of an agent must fit the expectations and habits of its users. Our re-design for the second focus day follows this idea: instead of requiring the developers to invoke skills, the agent uses the domain knowledge from the project context automatically.

Third, robotic engineering relies on visually observing how the robot moves in the simulated cell. Our agent only had access to the code, but not to the virtual station, which the developers also named as a problem. We believe that the resulting lack of spatial understanding contributes to the poor functionality of the generated code. A deeper integration of the agent with RobotStudio, e.g., through suitable MCP servers~\cite{anthropic2024mcp}, could help to overcome this limitation.

Fourth, we underestimated the impact of the terminal interface. While working in a terminal is common in software development, robotic engineers mostly work with graphical tools like RobotStudio and perceived the terminal as new and unfriendly on the first focus day. However, nobody named the usability as a problem after the second focus day, i.e., the developers quickly overcame this hurdle. Experiments with a high degree of control over a short period of time ignore such learning effects, even though they are an important confounding factor.

Finally, we note that the agent acted, to some degree, as an equalizer between the developers: the experience with robotic engineering did not determine the success with the agent, but the experience with AI did, i.e., something that is relatively new to all developers. The only task finished with the agent on the second focus day was solved by D2, who has the most AI experience, while D4, who rarely or never uses AI, did not finish any task with the agent and was the most disappointed by it. However, all developers in our case study are senior engineers with at least seven years of experience. Thus, we cannot generalize this observation to less experienced developers.

\begin{rqanswer}{Answer to RQ2}
The developers used the agent like a chat assistant and designing the agent to not require that developers manually invoking specific skills improved performance. Still, the engineers delegated all work, including trivial changes, which reduced efficiency and model performance. The experience with AI was more important for the successful use of the agent than the experience with robotic engineering.
\end{rqanswer}

\subsection{Recommendations for practitioners}\label{sec:disc-practice}

An on-premise agent with an open-weight model is technically feasible for SMEs. However, it is not a plug-and-play solution that directly increases productivity. Based on our case study, we recommend the following to SMEs that consider adopting agents for robotic engineering.

\emph{Train the usage, not only the tool.} Developers need to learn when to use the agent, when to change code manually, and when to start a new conversation. In our case, a short introduction to the tool was not sufficient. Instead, SMEs should plan dedicated training sessions with exercises in which the developers learn how to use the agent effectively, as well as time for the developers to get used to the agent in their daily work.

\emph{Start with suitable tasks.} Repetitive and well-specified generation tasks, e.g., trajectory matrices or drafts of documentation, are a better starting point than tasks that require a spatial understanding of the robot cell. This way, the developers can gain experience with the agent on tasks for which it already provides a benefit.

\emph{Configure the context.} The reference manual and the programming standards should be available to the agent automatically, e.g., through a project context, instead of relying on the developers to invoke skills. In our case, this improved the adherence to the programming standards and removed the missing ABB documentation from the problems named by the developers.

\subsection{Implications for researchers}\label{sec:disc-research}

Our case study also raises questions for research on agents in robotic engineering and, more generally, in specialized engineering domains.

\emph{Evaluate with developers in the loop.} Benchmarks on isolated tasks do not capture how developers actually use agents. Our results demonstrate this: while our prior work showed that an LLM can perform key steps of the tasks~\cite{10.1145/3786583.3786869}, the developers did not benefit from this with the agent. Thus, researchers should complement benchmarks with studies of real tasks in real contexts.

\emph{Study adoption over time.} Our two focus days provide only two snapshots, but already show strong learning effects, e.g., regarding the terminal interface. Longitudinal studies could show how long the learning curve is, whether the usage patterns change with more experience, and whether the productivity eventually exceeds the productivity without AI.

\emph{Report the configuration of agents.} Our results show that seemingly small differences in the configuration, i.e., skills that developers invoke vs.\ a project context that the agent uses automatically, can affect the outcomes. Thus, researchers should report such details to allow for comparisons between studies.

\section{Threats to validity}\label{sec:threats}

The main threat to the validity of our work is the small sample. Our results are based on four developers on the first and three on the second focus day, one company, one agent, and one model. Thus, our results do not generalize statistically. We mitigate this through a case study design that collects rich quantitative and qualitative data in a real industrial context~\cite{wohlin2012experimentation}. Still, other models, agents, or companies may lead to different results.

The team lead was the only assessor and the assessment was not blind, i.e., the team lead knew which solutions were created with the agent. To mitigate this threat, we complement the assessment with the validator, which checks the adherence to the programming standards independently of the team lead. Moreover, the time limits mean that we cannot measure how long unfinished tasks would have taken.

While our design has aspects of a controlled experiment, it is not a controlled experiment, because of both the sample size and the changes we applied between the focus days, like the re-design of the agent. We borrowed this methodology to structure the focus days and allow for a better comparison of results, not to establish causal claims. 

\section{Conclusion}
We report a case study with AKE Technologies, in which four experienced robotic engineers solved six tasks from real projects with and without an on-premise AI agent on two focus days. Our results show that the agent did not make the developers more productive: with the agent, they finished only 4 of 21 tasks within the time limit, in comparison to 14 of 21 without AI, and the functionality of their solutions was worse. The only exception is the generation of highly repetitive code. However, the solutions with the agent improved between the focus days, e.g., 5 of 8 instead of 2 of 10 solutions adhered to AKE's programming standards. We believe that this is due to both the re-design of the agent and the learning curve of the developers, whose success with the agent depended more on their experience with AI than on their experience with robotic engineering. Thus, agents are not a plug-and-play solution for robotic engineering in SMEs: they require training, a configuration that fits the domain, and time. In future work, we want to study how the productivity evolves over longer periods of time and whether a deeper integration of the agent with RobotStudio, i.e., access to the virtual station, overcomes the missing spatial understanding of the agent.

\bibliographystyle{Paper_2_IEEE_Format/bibtex/IEEEtran}
\bibliography{Paper_2_IEEE_Format/bibtex/Bibliography.bib}

\end{document}